# Superconducting and Resistive Tilted Coil Magnets

Andrey M. Akhmeteli and Andrew V. Gavrilin

*Abstract*—The mathematical foundation is laid for a relatively new type of magnets generating uniform transverse field – tilted coil magnets. These consist of concentric nested solenoidal coils with elliptical turns tilted at a certain angle to the central axis and current flowing in opposite directions in the coils tilted at opposite angles, generating a perfectly uniform transverse field. Both superconducting wire-wound and resistive Bitter tilted coils are discussed. An original analytical method is used to prove that the wire-wound tilted coils have the ideal distribution of the axial linear current density – "cosine-theta". Magnetic fields are calculated for a tilted Bitter coil magnet using an original exact solution for current density in an elliptical Bitter disk. Superconducting wire-wound tilted coil magnets may become an alternative for traditional dipole magnets for accelerators, and Bitter tilted coil magnets are attractive for rotation experiments with a large access port perpendicular to the field.

*Index Terms*—Transverse magnetic field, superconducting solenoids, Bitter magnets, dipole magnet.

## I. Introduction

THE idea to use superconducting wire-wound and resistive Bitter concentric nested tilted coils for uniform transverse field generation has been around for some time already (see [1] and references there) and has been considered in a number of recent publications [1-4]. Nevertheless, this type of coils can be still regarded as a novel one. In a "tilted coil magnet", elliptical turns are tilted at a certain angle ("tilt angle") to the central axis, and current is flowing in opposite directions in the coils tilted at opposite angles, generating a perfectly uniform transverse field, as longitudinal components of the field cancel each other out, leaving only a dipole component of the field (Figs. 1,2), if the average current density and coils' length are properly chosen and adjusted. The resultant transverse field is extremely uniform indeed, as calculations show [1-4], and tilted coil magnets, both superconducting [2] and resistive [1,2], look rather attractive from the practical point of view. Superconducting wire-wound tilted coil magnets may become an alternative for traditional dipole magnets for accelerators [2,3], and Bitter tilted coil magnets are attractive for rotation experiments with a large access port perpendicular to the field [5]. It is also noteworthy that a hybrid configuration, comprising a Bitter tilted coil insert and a superconducting tilted coil outsert, seems to be the only option to obtain uniform transverse magnetic field about 30T DC within a reasonably large volume of space.

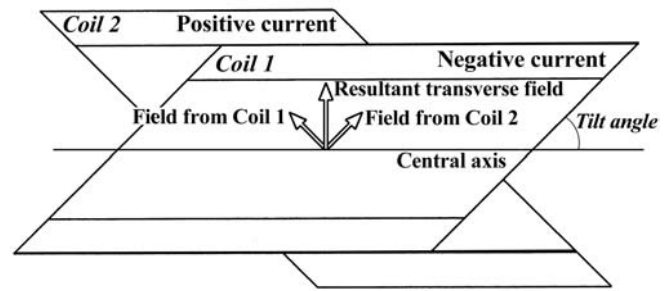

Fig. 1. Sketch of the simplest tilted coil magnet with two-coil configuration.

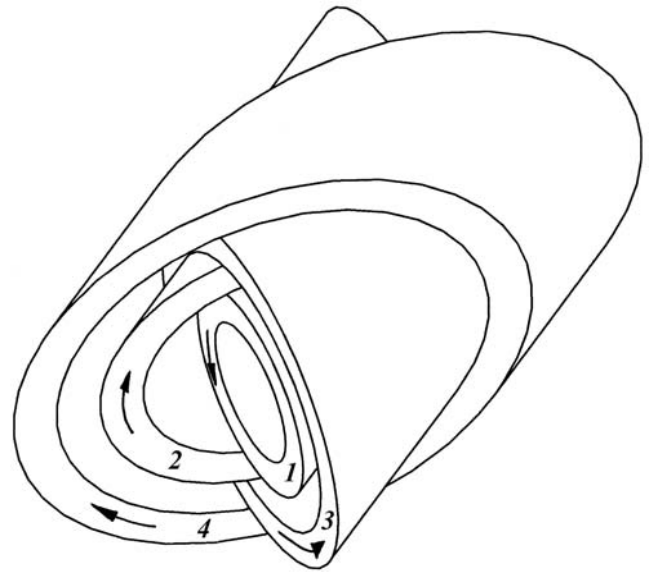

Fig. 2. 3D sketch of multiple tilted coil magnet configuration. Arrows show the current direction. The coils can be wire-wound multi-layer [2], wire-wound one-layer [3] and/or Bitter type ones [1].

## II. Wire-wound Tilted Coil as A Generalized "Cosine-theta" Configuration

Perhaps there exist several ways to explain mathematically why transverse field generated by a wire-wound tilted coil is so uniform. Still this problem has not been properly treated in the literature; at least, we failed to supply this paper with any references. Here we suggest a rather simple method to prove that wire-wound circular tilted coils have a generalized "cosine-theta" distribution of axial current density. In order to do so, it is sufficient to consider a fragment of one layer of the winding (Fig. 3).

Let us assume that the wire is wound in such a way that centers of cross-sections of the wire lie on the following parametric curve (a tilted cylindrical helix, Fig. 3):

$$\mathbf{P}(\theta) = \{R\cos\theta, R\sin\theta, Rq\sin\theta + w\theta\}.$$

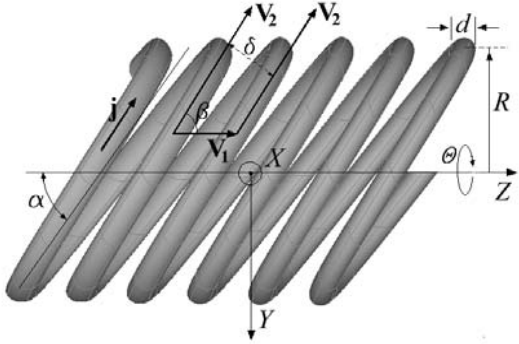

Fig. 3. Sketch of a fragment of one layer of a tilted coil wound with a round conductor (wire).

An increase of $\theta$ by $2\pi$ leads to a shift by a vector (independent of $\theta$): $\mathbf{V}_1 = \{0, 0, 2\pi w\}$. A vector tangent to curve $\mathbf{P}(\theta)$ may be found by differentiation with respect to $\theta$:

$$\mathbf{V}_2 = \mathbf{V}_2(\theta) = \{-R\sin\theta, R\cos\theta, Rq\cos\theta + w\},$$

so the current vector equals

$$\mathbf{I} = \mathbf{I}(\theta) = I \cdot \mathbf{V}_2 / V_2,$$

where $I = |\mathbf{I}(\theta)| = const$ is the current in the wire, $V_1 = |\mathbf{V}_1| = const$, $V_2 = |\mathbf{V}_2|$ are the magnitudes of the relevant vectors. The average current density vector equals

$$\mathbf{j} = \mathbf{j}(\theta) = \mathbf{I}/(d\delta),$$

where $d$ is the diameter of the wire, $\delta = \delta(\theta)$ is the distance between tangents to curve $\mathbf{P}(\theta)$ in two points where parameter $\theta$ differs by $2\pi$ (Fig. 3).

It should be noted that the magnitude of the average current density $j = |\mathbf{j}|$ depends on $\theta$, although neither the current, $I$, nor the wire cross-section depend on $\theta$. This is due to the fact that the winding cannot be uniformly tight in this geometry, and the gap, $g$, between two adjacent turns of the winding depends on $\theta$: $g(\theta) \approx \delta(\theta) - d$. As can be seen, the gap is minimal at $\theta \approx k\pi$, where $k$ is integer; in particular, in the vicinity of these points the adjacent turns can touch each other if the winding is closely packed; the gap is widest at $\theta = \pi/2 + k\pi$ and depends on the tilt angle $\alpha$ (Fig. 3).

Evidently,

$$\delta = V_1 \sin\beta = |\mathbf{V}_1 \times \mathbf{V}_2|/V_2 = 2\pi wR/V_2, \quad \beta = \beta(\theta),$$

therefore

$$\mathbf{j} = I \cdot \mathbf{V}_2/(V_2 d\delta) = I \cdot \mathbf{V}_2/(2\pi wRd),$$

and the axial component of the average current density turns out to be

$$j_Z = I\frac{Rq\cos\theta + w}{2\pi wRd} = j_{Z\max} \cdot \cos\theta + j_Z^{const}.$$

It is noteworthy that a constant component of the axial current appears due to the helical path of wire – it disappears if each turn is considered as a flat ellipse. Formally speaking, the presence of the constant component of axial current is a distinction, compared to classic cosine-theta coils, which basically do not have such a "makeweight". Nevertheless, this is "a distinction without a difference", which does not deteriorate the ideal uniformity of transverse field within the bore.

It is easy to see how the parameters of curve $\mathbf{P}(\theta)$ are related to the parameters of the one-layer winding: the external and the internal radii equal $R + d/2$ and $R - d/2$, respectively; $q = \cot\alpha$. Parameter $w$ may be found from the following equation:

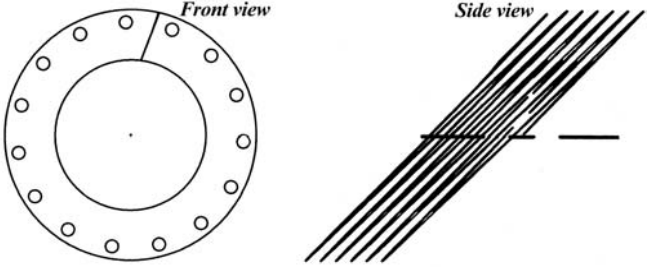

Fig. 4. Fragment of a tilted Bitter coil with elliptical disks.

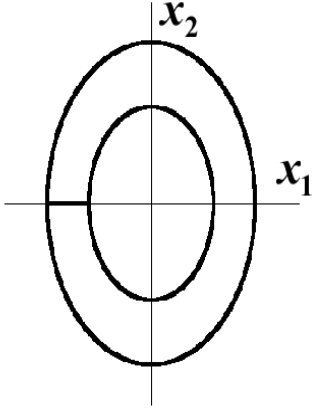

Fig. 5. Sketch of an elliptical disk with a coordinate system aligned with it.

$$d \approx \delta(0) = \min(\delta(\theta)) = \min\left(\frac{2\pi wR}{V_2}\right) =$$

$$\min\left(\frac{2\pi wR}{\sqrt{R^2 + (Rq\cos(\theta)+w)^2}}\right) = \frac{2\pi wR}{\sqrt{R^2 + (Rq+w)^2}},$$

assuming that the coil is wound as tightly as possible.

Obviously, the more layers of winding are used, the higher transverse field can be obtained at the same current. As to the axial component of the magnetic field, also generated by a tilted coil, it can be easily cancelled out through the use of two nested oppositely tilted coils energized with opposite polarity [2,3] (Fig. 1). The same is correct for Bitter tilted magnets discussed below.

### III. RESISTIVE BITTER TILTED COILS

As opposed to wire-wound tilted coils, Bitter tilted coils (Fig. 4) are much more complicated for analysis. The magnetic field from a tilted coil can be adequately approximated by the field of a number of identical elliptical disks parallel to each other. In our analysis, we neglect the presence of numerous holes in the disks.

To calculate the magnetic field of a tilted coil, let us calculate the current density distribution in one elliptical Bitter disk. While for a conventional circular Bitter disk the current density distribution obeys a very simple analytical formula [6], this is not the case for the current density in an elliptical disk (Fig. 5). However, a more complex exact solution was found for the latter for an arbitrary tilt angle.

Let us consider an elliptical disk defined by the following equations in Cartesian coordinates $x_1$, $x_2$, $x_3$ (Fig. 5):

$$p_1^2 \le wx_1^2 + x_2^2 \le p_2^2,$$

$$-\frac{d}{2} \le x_3 \le \frac{d}{2}.$$

Here $w > 1$.

We assume that the disk is cut in the half plane $x_1 < 0$, and static electric potentials $\Phi_1$ and $\Phi_2$ are applied to the surfaces of the cut. Due to the symmetry of the problem, the potential (and, consequently, the current density) does not depend on $x_3$ (we assume that the conductivity $\sigma$ is constant within the disk). As the current density $\mathbf{j}(x_1, x_2)$ is static, the continuity equation has the following form:

$$\nabla \mathbf{j}(x_1, x_2) = 0.$$

As

$$-\nabla \Phi(x_1, x_2) = \mathbf{E}(x_1, x_2) = \mathbf{j}(x_1, x_2)/\sigma,$$

where $\mathbf{E}(x_1, x_2)$ is the electric field strength, the problem may be stated as follows.

We seek the partial solution $\Phi(x_1, x_2)$ of the equation $\Delta\Phi = 0$ in the area limited by two homothetic ellipses that are defined by the equations:

$$wx_1^2 + x_2^2 = p_1^2,$$

$$wx_1^2 + x_2^2 = p_2^2,$$

with a cut along the abscissa axis (in the left half-plane).

On the upper and the lower banks of the cut, the function equals $\pi$ and $-\pi$, correspondingly (these values were chosen for convenience; if necessary, this solution may be modified in an obvious way to accommodate for the actual values of $\Phi_1$ and $\Phi_2$). At the elliptic boundaries, the projection of the gradient of this function on the normal to the ellipses equals zero (as there is no charge flow through these boundaries).

Using a standard procedure (see, e.g., [7]), one can prove that the function sought realizes an extremum of the functional

$$J(\Phi) = \int_S dS (\nabla \Phi)^2,$$

where $S$ is the area between the ellipses, on the set of functions that have the same values on the banks of the cut ($\pm \pi$).

The equation $\Delta \Phi = 0$ may be solved by separation of variables in non-orthogonal coordinates (cf. [8]):

$$u = \ln r + \frac{1}{2}\ln((w-1)\cos(2\varphi) + w + 1),$$

$$v = \varphi,$$

where $r$ and $\varphi$ are the polar coordinates. This method is equivalent to solving the equation by the Ritz method (see, e.g., [7]) with the following functions (multipoles) used as the basis:

$$(\alpha \ln r + \beta)(\gamma \varphi + \delta),$$

$$Cr^{\pm n} \exp(\pm in\varphi),$$

where $\alpha, \beta, \gamma, \delta, C$ are constants. The sought-for solution $\Phi$ will be a linear combination of these solutions. In view of the boundary conditions on the banks of the cut, the solution may be sought in the form:

$$\Phi(r,\varphi) = \varphi + \sum_{n \neq 0} c_n r^n \sin(n\varphi),$$

where $n$ is integer (possibly, negative).

The coefficients $c_n$ are evaluated from the condition of extremum of the functional. In polar coordinates, the gradient has the following form:

$$\nabla \Psi = \frac{\partial \Psi}{\partial r}\mathbf{i}_r + \frac{1}{r}\frac{\partial \Psi}{\partial \varphi}\mathbf{i}_\varphi,$$

$$\nabla \varphi = \frac{1}{r}\mathbf{i}_\varphi,$$

$$\nabla(r^n \sin(n\varphi)) = nr^{n-1}\sin(n\varphi)\mathbf{i}_r + r^{n-1}n\cos(n\varphi)\mathbf{i}_\varphi,$$

$$\int dS(\nabla \Phi)^2 = \int dS(\nabla(\varphi + \sum_{n \neq 0} c_n r^n \sin(n\varphi)))^2.$$

We may write the following:

$$\nabla \Psi = \sum_n g_n r^{n-1}(\sin(n\varphi)\mathbf{i}_r + \cos(n\varphi)\mathbf{i}_\varphi),$$

where

$$g_n = \begin{cases} 1 & n = 0 \\ nc_n & n \neq 0 \end{cases}.$$

Then

$$(\nabla \Psi)^2 = \sum_{m,n} g_n g_m r^{m+n-2}(\sin(n\varphi)\sin(m\varphi) + \cos(n\varphi)\cos(m\varphi)) = \sum_{m,n} g_n g_m r^{m+n-2}\cos((m-n)\varphi),$$

$$\int_S (\nabla \Psi)^2 dS = \sum_{m,n} g_n g_m A_{nm},$$

$$A_{nm} = \int_S r^{m+n-2}\cos((m-n)\varphi) dS = \int_{-\pi}^{\pi} d\varphi \cos((m-n)\varphi) \int_{r_1(\varphi)}^{r_2(\varphi)} rdr \cdot r^{m+n-2} =$$

$$= \begin{cases} \int\limits_{-\pi}^{\pi} d\varphi \cos((m-n)\varphi)(\ln r_2(\varphi) - \ln r_1(\varphi)) & (m+n=0) \\ \int\limits_{-\pi}^{\pi} d\varphi \cos((m-n)\varphi) \frac{r_2^{m+n}(\varphi) - r_1^{m+n}(\varphi)}{m+n} & (m+n \neq 0) \end{cases},$$

where $r_2(\varphi)$ and $r_1(\varphi)$ are the coordinates $r$ of the points of the larger and the lesser ellipses with coordinate $\varphi$, correspondingly.

The equation of the larger ellipse is:
$$wx_1^2 + x_2^2 = p_2^2$$
or
$$wr^2 \cos^2 \varphi + r^2 \sin^2 \varphi = p_2^2,$$

$$r_2^2(\varphi) = \frac{p_2^2}{w\cos^2\varphi + \sin^2\varphi} = \frac{p_2^2}{1+(w-1)\cos^2\varphi} = \frac{p_2^2}{1+(w-1)\frac{\cos(2\varphi)+1}{2}} = \frac{2p_2^2}{(w-1)\cos(2\varphi)+w+1}.$$

Similarly,
$$r_1^2(\varphi) = \frac{2p_1^2}{(w-1)\cos(2\varphi)+w+1}.$$

Thus, if $m+n=0$, then
$$A_{nm} = \int\limits_{-\pi}^{\pi} d\varphi \cos((m-n)\varphi) \cdot \frac{1}{2}\left(\ln \frac{2p_2^2}{(w-1)\cos(2\varphi)+w+1} - \ln \frac{2p_1^2}{(w-1)\cos(2\varphi)+w+1}\right) =$$

$$= \int\limits_{-\pi}^{\pi} d\varphi \cos((m-n)\varphi) \cdot \ln \frac{p_2}{p_1} = \delta_{mn} \cdot 2\pi \cdot \ln \frac{p_2}{p_1},$$

where $\delta_{mn}$ is the Kronecker's delta.

If $m+n \neq 0$, then
$$A_{nm} = \int\limits_{-\pi}^{\pi} d\varphi \cos((m-n)\varphi) \cdot \frac{1}{m+n}\left(\left(\frac{2p_2^2}{(w-1)\cos(2\varphi)+w+1}\right)^{\frac{m+n}{2}} - \left(\frac{2p_1^2}{(w-1)\cos(2\varphi)+w+1}\right)^{\frac{m+n}{2}}\right) =$$

$$= \frac{2^{\frac{m+n}{2}}}{m+n}(p_2^{m+n} - p_1^{m+n}) \int\limits_{-\pi}^{\pi} d\varphi \cos((m-n)\varphi)((w-1)\cos(2\varphi)+w+1)^{-\frac{m+n}{2}}.$$

If we only leave harmonics with numbers from $-k$ to $k$, then, differentiating quadratic form $\sum\limits_{n,m=-k}^{k} A_{nm} g_n g_m$ with respect to $g_n$ ($n \neq 0$) and equating the results to zero, we obtain a system of linear equations that determines $g_n$ ($n \neq 0$).

Let us consider the integral
$$\int\limits_{-\pi}^{\pi} d\varphi \cos((m-n)\varphi)((w-1)\cos(2\varphi)+w+1)^{-\frac{m+n}{2}}.$$

If $(m-n)$ is odd, this integral equals zero: the function $((w-1)\cos(2\varphi)+w+1)^{-\frac{m+n}{2}}$ is periodic with period $\pi$, so there are even harmonics only ($\exp(\pm 2ik\varphi)$, where $k$ is integer) in its Fourier series, and the integral equals the coefficient of an odd harmonic to a factor. Therefore, in the series for $\Phi$ we may leave even terms only:
$$\Phi(r,\varphi) = \varphi + \sum\limits_{k \neq 0} c_{2k} r^{2k} \sin(2k\varphi),$$
where $k$ is integer.

Let $m = 2l$, $n = 2k$; the integral has the following form:

$$\int_{-\pi}^{\pi} d\varphi \cos(2(l-k)\varphi)((w-1)\cos(2\varphi)+w+1)^{-l-k} = \frac{1}{2}\int_{-2\pi}^{2\pi} du \cos((l-k)u)((w-1)\cos(u)+w+1)^{-l-k} =$$

$$= \int_{-\pi}^{\pi} du \cos((l-k)u)((w-1)\cos(u)+w+1)^{-l-k}.$$

So, let us evaluate

$$I = \int_{-\pi}^{\pi} du \cos((l-k)u)(\cos(u)+q)^{-l-k} = \text{Re} \int_{-\pi}^{\pi} du \exp(i(l-k)u)(\cos(u)+q)^{-l-k},$$

where $q = \dfrac{w+1}{w-1} > 0$.

Let $z = \rho \exp(iu)$. On the unit circle, $dz = \rho i \exp(iu) du = i \exp(iu) du$.

$$I_1 = \int_{-\pi}^{\pi} du \exp(i(l-k)u)(\cos(u)+q)^{-l-k} = \oint \frac{dz}{i \exp(iu)} \cdot \exp(i(l-k)u) \cdot (\cos(u)+q)^{-l-k} =$$

$$= \oint dz (-i) z^{l-k-1} \left( \frac{z+z^{-1}}{2} + q \right)^{-l-k} = -i \cdot 2^{l+k} \oint dz \cdot z^{2l-1} (z^2 + 2qz + 1)^{-l-k}.$$

The integration contour is the unit circle. Inside the unit circle, there may be poles in the points $z = 0$ and $z = -q + \sqrt{q^2 - 1}$.

Let us first consider the case where $l + k < 0$. Let us evaluate the integral

$$\oint dz \cdot z^m (z^2 + 2qz + 1)^n,$$

where $n > 0$. If in the same time $m \geq 0$, there are no singular points inside the unit circle, so the integral equals zero. Let consider the integral

$$\oint dz \cdot z^{-m} (z^2 + 2qz + 1)^n,$$

where $n, m > 0$.

$$(z^2 + 2qz + 1)^n = ((z+q)^2 - (q^2 - 1))^n = \sum_{j=0}^{n} \binom{n}{j} (z+q)^{2j} (-(q^2 - 1))^{n-j},$$

where $\dbinom{n}{j} = \dfrac{n!}{j!(n-j)!}$ is the number of all combinations of $n$ elements taken $j$ at a time.

In this connection, let us evaluate the integral

$$\oint dz \cdot z^{-m} (z+q)^k,$$

where $m > 0$, $k \geq 0$. It has only one pole of order $m$ in the zero point. If inside the integration contour the only singular points are poles, integral $\oint_C dz \cdot f(z)$ equals $2\pi i \sum_k \text{Res}\, f(z_k)$, where $z_k$ are poles, and $\text{Res}\, f(z_k)$ is the residue in the pole $z_k$. A residue in the point $a$ that is a pole of order $m$ equals

$$\text{Res}\, f(a) = \frac{1}{(m-1)!} \lim_{z \to a} \frac{d^{m-1}}{dz^{m-1}} ((z-a)^m f(z)).$$

In our case the residue in zero equals

$$\frac{1}{(m-1)!} \lim_{z \to 0} \frac{d^{m-1}}{dz^{m-1}} (z+q)^k.$$

If $k < m-1$, the residue equals zero, and if $k \geq m-1$, it equals

$$\frac{1}{(m-1)!}k(k-1)\ldots(k-m+2)\cdot q^{k-m+1} = \frac{1}{(m-1)!}\frac{k!}{(k-m+1)!}\cdot q^{k-m+1} = \binom{k}{m-1}\cdot q^{k-m+1}.$$

Thus, the evaluation of the integral $\oint dz \cdot z^m (z^2 + 2qz + 1)^n$, where $n > 0$, is completed.

Now let us consider the integral

$$\oint dz \cdot z^m (z^2 + 2qz + 1)^{-n},$$

where $n > 0$.

$$z^2 + 2qz + 1 = \left(z + q + \sqrt{q^2 - 1}\right)\left(z + q - \sqrt{q^2 - 1}\right) = -\left(z + q + \sqrt{q^2 - 1}\right)\left(\sqrt{q^2 - 1} - z - q\right) =$$

$$= -2\sqrt{q^2 - 1}\left(\frac{1}{2} + \frac{z+q}{2\sqrt{q^2 - 1}}\right) \cdot 2\sqrt{q^2 - 1}\left(\frac{1}{2} - \frac{z+q}{2\sqrt{q^2 - 1}}\right).$$

Then

$$\frac{1}{z^2 + 2qz + 1} = \frac{-1}{4(q^2 - 1)\left(\frac{1}{2} + \frac{z+q}{2\sqrt{q^2 - 1}}\right)\left(\frac{1}{2} - \frac{z+q}{2\sqrt{q^2 - 1}}\right)} = -\frac{1}{4(q^2 - 1)}ab,$$

where

$$a = \frac{1}{\frac{1}{2} + \frac{z+q}{2\sqrt{q^2 - 1}}}, \quad b = \frac{1}{\frac{1}{2} - \frac{z+q}{2\sqrt{q^2 - 1}}}.$$

Note that $a + b = ab$.

Let us prove that if $a + b = ab$ and $n > 0$, then

$$(ab)^n = \sum_{i=1}^{n} c_i^{[n]}(a^i + b^i),$$

where $c_i^{[n]}$ is defined by the following recurrence formulae:

$c_1^{[1]} = 1$,

$c_i^{[n]} = (1 + \delta_{i1}) \sum_{j=i-1+\delta_{i1}}^{n-1} c_j^{[n-1]}$ for $n > 1$ ($\delta_{ij}$ is the Kronecker's delta).

Proof.

It is evident that the statement is true for $n = 1$. Let us assume that the statement is true for any integer $k$ such that $1 \leq k \leq n - 1$ ($n > 1$) and prove that it is true for $n$.

As $a + b = ab$, we obtain $b = \frac{a}{a-1}$, $a = \frac{b}{b-1}$.

Then

$$(a+b)(a^i + b^i) = a^{i+1} + ba^i + ab^i + b^{i+1} = a^{i+1} + \frac{a}{a-1}a^i + \frac{b}{b-1}b^i + b^{i+1} =$$

$$= a^{i+1} + \frac{a^{i+1}}{a-1} + \frac{b^{i+1}}{b-1} + b^{i+1} = \frac{a^{i+2}}{a-1} + \frac{b^{i+2}}{b-1} = \frac{a^{i+2} - a + a}{a-1} + \frac{b^{i+2} - b + b}{b-1} =$$

$$= a(1 + a + \ldots + a^i) + b + b(1 + b + \ldots + b^i) + a =$$

$$(a + a^2 + \ldots + a^{i+1}) + b + (b + b^2 + \ldots + b^{i+1}) + a = \sum_{k=1}^{i+1}(a^k + b^k) + a + b.$$

$$a^n b^n = \sum_{i=1}^{n} c_i^{[n]}(a^i + b^i) = (a+b)\sum_{i=1}^{n-1} c_i^{[n-1]}(a^i + b^i) =$$

$$= \sum_{i=1}^{n-1} c_i^{[n-1]} \left( \sum_{k=1}^{i+1} (a^k + b^k) + a + b \right) = \text{(we change the order of summation)} =$$

$$= \left( \sum_{k=1}^{n} (a^k + b^k) \sum_{i=\max(1,k-1)}^{n-1} c_i^{[n-1]} \right) + \sum_{i=1}^{n-1} c_i^{[n-1]} (a+b).$$

Hence,

$$c_i^{[n]} = \sum_{j=\max(1,i-1)}^{n-1} c_j^{[n-1]} + \delta_{i1} \sum_{j=1}^{n-1} c_j^{[n-1]} \quad (n > 1),$$

or

$$c_i^{[n]} = (1 + \delta_{i1}) \sum_{j=\max(1,i-1)}^{n-1} c_j^{[n-1]} = (1 + \delta_{i1}) \sum_{j=i-1+\delta_{i1}}^{n-1} c_j^{[n-1]}.$$

This completes the proof of the statement.

Thus, for $n > 0$

$$\oint dz \cdot z^m (z^2 + 2qz + 1)^{-n} =$$

$$= \oint dz \cdot z^m \left( -\frac{1}{4(q^2-1)} \right)^n \frac{1}{\left( \frac{1}{2} + \frac{z+q}{2\sqrt{q^2-1}} \right)^n} \frac{1}{\left( \frac{1}{2} - \frac{z+q}{2\sqrt{q^2-1}} \right)^n} =$$

$$= \oint dz \cdot z^m \left( -\frac{1}{4(q^2-1)} \right)^n \sum_{i=1}^{n} c_i^{[n]} \left( \left( \frac{1}{\frac{1}{2} + \frac{z+q}{2\sqrt{q^2-1}}} \right)^i + \left( \frac{1}{\frac{1}{2} - \frac{z+q}{2\sqrt{q^2-1}}} \right)^i \right) =$$

$$= \left( -\frac{1}{4(q^2-1)} \right)^n \sum_{i=1}^{n} c_i^{[n]} \left( 2\sqrt{q^2-1} \right)^{(i)} \oint dz \cdot z^m \left( \left( \frac{1}{z+q+\sqrt{q^2-1}} \right)^i + (-1)^i \left( \frac{1}{z+q-\sqrt{q^2-1}} \right)^i \right).$$

Therefore, it is sufficient to evaluate the integrals

$$\oint dz \cdot z^m \frac{1}{\left( z + q \pm \sqrt{q^2-1} \right)^n} \quad (n>0).$$

Let us first evaluate the integral

$$\oint dz \cdot z^m \frac{1}{\left( z + q + \sqrt{q^2-1} \right)^n}.$$

It may only have a pole in zero. The relevant residue equals zero if $m \geq 0$. Let us evaluate the residue in zero for the integral

$$\oint dz \cdot z^{-m} \frac{1}{\left( z + q + \sqrt{q^2-1} \right)^n} \quad (m > 0), \text{ it equals}$$

$$\frac{1}{(m-1)!} \lim_{z \to 0} \frac{d^{m-1}}{dz^{m-1}} \left( z + q + \sqrt{q^2-1} \right)^{-n} = \frac{1}{(m-1)!} (-n)(-n-1)\ldots(-n-m+2) \left( q + \sqrt{q^2-1} \right)^{-n-m+1} =$$

$$= \frac{1}{(m-1)!} (-1)^{m-1} \frac{(n+m-2)!}{(n-1)!} \left( q + \sqrt{q^2-1} \right)^{-n-m+1}.$$

Now let us evaluate the integral

$$\oint dz \cdot z^m \frac{1}{\left(z+q-\sqrt{q^2-1}\right)^n} \quad (n>0).$$

It may have poles in zero and in the point $z = -q + \sqrt{q^2-1}$. The residue in zero is evaluated as above, i.e. it equals zero if $m \geq 0$, and the residue in zero for the integral $\oint dz \cdot z^{-m} \frac{1}{\left(z+q-\sqrt{q^2-1}\right)^n}$ ($m > 0$) equals

$$\frac{1}{(m-1)!} \lim_{z \to 0} \frac{d^{m-1}}{dz^{m-1}} \left(z+q-\sqrt{q^2-1}\right)^{-n} = \frac{1}{(m-1)!}(-n)(-n-1)\ldots(-n-m+2)\left(q-\sqrt{q^2-1}\right)^{-n-m+1} =$$

$$= \frac{1}{(m-1)!}(-1)^{m-1}\frac{(n+m-2)!}{(n-1)!}\left(q-\sqrt{q^2-1}\right)^{-n-m+1}.$$

Now let us evaluate the residue of the integral $\oint dz \cdot z^m \frac{1}{\left(z+q-\sqrt{q^2-1}\right)^n}$ in the point $z = -q + \sqrt{q^2-1}$. It equals

$$\frac{1}{(n-1)!} \lim_{z \to -q+\sqrt{q^2-1}} \frac{d^{n-1}}{dz^{n-1}} z^m = \frac{1}{(n-1)!}(m)(m-1)\ldots(m-n+2)\left(-q+\sqrt{q^2-1}\right)^{m-n+1} =$$

$$= \begin{cases} \frac{1}{(n-1)!} \frac{m!}{(m-n+1)!}\left(-q+\sqrt{q^2-1}\right)^{m-n+1} & (m \geq n-1) \\ 0 & (0 \leq m < n-1) \\ \frac{1}{(n-1)!} m(m-1)\ldots(m-n+2)\left(-q+\sqrt{q^2-1}\right)^{m-n+1} & (m < 0) \end{cases}$$

Thus, a complete analytical algorithm of evaluation of the integral $\oint dz \cdot z^{2l-1}(z^2+2qz+1)^{-l-k}$ may be developed.

The amplitudes of the harmonics may also be evaluated by minimization of a different quadratic form. The modified version of the algorithm is of significant interest as it provides a simple and natural estimate of the error for a finite number of harmonics. Let us evaluate a weighted line integral of the squared normal component of the potential gradient over the length of both boundary ellipses:

$$I_{\partial S} = \int_{\partial S} (\nabla \Psi)_n^2 \sqrt{\frac{1+2q\cos(2\varphi)+q^2}{\cos(2\varphi)+q}} dl.$$

It is evident that the weight varies between two positive values, as $q > 1$, so the integral gives an estimate of the error of the solution (the integral is zero for the exact solution, as there is no normal component of the current density at the elliptical boundary). If

$$\nabla \Psi = \sum_{k=-l}^{l} g_{2k} r^{2k-1}(\sin(2k\varphi)\mathbf{i}_r + \cos(2k\varphi)\mathbf{i}_\varphi),$$

one can show that the normal component of the potential gradient on the boundary ellipses equals

$$(\nabla \Psi)_n = \sum_{k=-l}^{l} g_{2k} r^{2k-1}(q\sin(2k\varphi)+\sin(2(k-1)\varphi))\frac{1}{\sqrt{1+2q\cos(2\varphi)+q^2}},$$

and

$$ds = \frac{\sqrt{1+2q\cos(2\varphi)+q^2}}{\cos(2\varphi)+q} r |d\varphi|.$$

Then one can show that

$$I_{\partial S} = \sum_{k,j=-l}^{l} g_{2k} g_{2j} D_{kj},$$

where

$$D_{kj} = \left(\frac{2}{w-1}\right)^{k+j-\frac{1}{2}} \left(p_2^{2k+2j-1} + p_1^{2k+2j-1}\right) \int_{-\pi}^{\pi} (\cos(2k\varphi) + q)^{-k-j-1} \cdot$$

$$\cdot (q\sin(2k\varphi) + \sin(2(k-1)\varphi)) \cdot (q\sin(2j\varphi) + \sin(2(j-1)\varphi)) d\varphi =$$

$$= \left(\frac{2}{w-1}\right)^{k+j-\frac{1}{2}} \left(p_2^{2k+2j-1} + p_1^{2k+2j-1}\right) I_{kj}.$$

Then one can obtain:

$$I_{kj} = \int_{-\pi}^{\pi} (\cos(u) + q)^{-k-j-1} (q\sin(ku) + \sin((k-1)u)) \cdot (q\sin(ju) + \sin((j-1)u)) du =$$

$$= \frac{1}{2}((q^2+1)I(q,k-j,-k-j-1) - q^2 I(q,k+j,-k-j-1) + qI(q,k-j-1,-k-j-1) -$$

$$- 2qI(q,k+j-1,-k-j-1) + qI(q,k-j+1,-k-j-1) - I(q,k+j-2,-k-j-1)),$$

where

$$I(q,m,n) = \int_{-\pi}^{\pi} \cos(mu)(\cos(u) + q)^n du.$$

The algorithm of evaluation of $I(q,m,n)$ is outlined above (we should just add that if $n = 0$, this integral is trivial and equals $2\pi\delta_{0m}$).

It should be noted that this modified version of the algorithm has another advantage: the matrix of the relevant system of linear equations is symmetric and positive definite, so the system may be solved using a faster and numerically more stable method (Cholesky decomposition). This is important, as for ellipses with higher eccentricity, numerical stability becomes a difficult issue, so multiple-precision arithmetic was used in this work. As more harmonics are left in the solution, the relevant series converges to the exact solution exponentially fast.

Summarizing, we may say that the exact solution for the electric potential (and, consequently, for the current density distribution) in the elliptical Bitter disk was found by the Ritz method with multipoles as the appropriate function basis. The solution can be naturally extended over several disks and allows easy differentiation.

For identical disks, it is sufficient to calculate the coefficients of the series just once. Current density $\mathbf{j}(\mathbf{r}_1)$ in an elliptical Bitter disk creates magnetic field in point $\mathbf{r}_2$ equal (up to a constant factor) to the following integral over the volume of the disk:

$$\mathbf{T} = \int_V \frac{\mathbf{j}(\mathbf{r}_1) \times \mathbf{R}}{R^3} d\mathbf{r}_1 = \int_V \left(\mathbf{j}(\mathbf{r}_1) \times \nabla_{r_1}\left(\frac{1}{R}\right)\right) d\mathbf{r}_1,$$

where $\mathbf{R} = \mathbf{r}_2 - \mathbf{r}_1$, $R = |\mathbf{R}|$. For a curl of a product of a scalar field and a vector field we have

$$\nabla_{r_1} \times \left(\mathbf{j}(\mathbf{r}_1)\frac{1}{R}\right) = \frac{1}{R}\left(\nabla_{r_1} \times \mathbf{j}(\mathbf{r}_1)\right) - \mathbf{j}(\mathbf{r}_1) \times \nabla_{r_1}\left(\frac{1}{R}\right) = -\mathbf{j}(\mathbf{r}_1) \times \nabla_{r_1}\left(\frac{1}{R}\right),$$

as $\nabla_{r_1} \times \mathbf{j}(\mathbf{r}_1) = 0$. The well-known formula for a volume integral of a curl yields

$$\mathbf{T} = -\int_V \nabla_{r_1} \times \left(\mathbf{j}(\mathbf{r}_1)\frac{1}{R}\right) d\mathbf{r}_1 = -\int_S \frac{1}{R}(d\mathbf{S} \times \mathbf{j}(\mathbf{r}_1)).$$

The elliptical disk was defined above in Cartesian coordinates $x_1$, $x_2$, $x_3$. As component $j_3$ equals zero, component $T_3$ equals the integral over the lateral surfaces of the disk. It may be shown that

$$T_3 = \left(\int_{S_i} - \int_{S_o}\right) \frac{1}{R} |\mathbf{j}(\mathbf{r}_1)| \operatorname{sgn}(j_\varphi(\mathbf{r}_1)) dl dx_3,$$

where $S_i$ is the inner lateral surface of the disk ($wx_1^2 + x_2^2 = p_1^2$, $-\frac{d}{2} \le x_3 \le \frac{d}{2}$), $S_o$ is the outer lateral surface of the disk ($wx_1^2 + x_2^2 = p_2^2$, $-\frac{d}{2} \le x_3 \le \frac{d}{2}$), $\text{sgn}(j_\varphi(\mathbf{r}_1))$ is the sign of the azimuthal component of the current density in polar coordinates, $dl = \sqrt{dx_1^2 + dx_2^2}$ is the element of length of the ellipse ($wx_1^2 + x_2^2 = p_1^2$ and $wx_1^2 + x_2^2 = p_2^2$, correspondingly). It should be taken into account that $j_3(\mathbf{r}_1) = 0$, the normal component of the current density vanishes on the surface of the disk, and the current density does not depend on $x_3$, so $\mathbf{j}(\mathbf{r}_1) = \mathbf{j}(x_1, x_2)$. Therefore, it is not difficult to perform integration with respect to $x_3$, obtaining

$$T_3 = \left(\int_{C_i} - \int_{C_o}\right) |\mathbf{j}(\mathbf{r}_1)| \text{sgn}(j_\varphi(\mathbf{r}_1)) \frac{\ln\left|y_3 + \frac{d}{2} + \sqrt{\left(y_3 + \frac{d}{2}\right)^2 + (y_1 - x_1)^2 + (y_2 - x_2)^2}\right|}{\ln\left|y_3 - \frac{d}{2} + \sqrt{\left(y_3 - \frac{d}{2}\right)^2 + (y_1 - x_1)^2 + (y_2 - x_2)^2}\right|} dl,$$

where $C_i$ and $C_o$ are the ellipses $wx_1^2 + x_2^2 = p_1^2$ and $wx_1^2 + x_2^2 = p_2^2$, correspondingly, and $y_1$, $y_2$, $y_3$ are the Cartesian coordinates of $\mathbf{r}_2$ ($\mathbf{r}_2 = \{y_1, y_2, y_3\}$).

The integral over the bases of the disk yields the other two components of the field ($T_1$ and $T_2$):

$$\{T_1, T_2\} = \left(\int_{S_t} - \int_{S_b}\right) \frac{1}{R} \{j_2, -j_1\} dS,$$

where $S_t$ is the top base of the disk ($p_1^2 \le wx_1^2 + x_2^2 \le p_2^2$, $x_3 = \frac{d}{2}$), $S_b$ is the bottom base of the disk ($p_1^2 \le wx_1^2 + x_2^2 \le p_2^2$, $x_3 = -\frac{d}{2}$). As the solution for current density is presented as a series in multipoles, the integral over the bases of the disk can be reduced to a sum of one-dimensional integrals (after integration with respect to $r$ in the polar system of coordinates), but this was not done in this work, and the two-dimensional integral was calculated.

It should be mentioned that so far we cannot explain as easily as we did for wire-wound coils the extremely high uniformity of the field that can be obtained with tilted Bitters, as our numerical experiments clearly show [1] – the solution for current density in the disk is too complicated as opposed to that for wire-wound tilted coils.

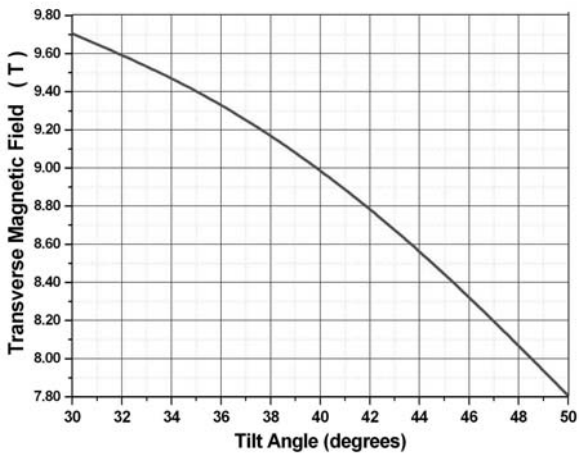

Fig. 6. The transverse component of magnetic field on the axis of a single long tilted Bitter coil, in the central part of the coil, *vs* the tilt angle. The axial component of the magnetic field is not shown.

IV. TILTING EFFECT

Good examples of uniform transverse magnetic field that can be obtained with tilted coils are given in previous publications [1-4]. The effect of tilt angle on the transverse field strength was also discussed in literature [3], albeit rather briefly and for wire-wound coils only. In order to give a flavor of this effect for tilted Bitter coils, we conducted some calculations for a particular single Bitter coil with rather typical parameters: 25 mm inner diam, 85 mm outer diam, 2 mm disk (turn) thickness, a total of 300 disks (turns), 40 kA transport current. In calculations, we varied the tilt angle only, keeping other parameters constant. Despite the fact that the coil's length somewhat changes as the tilt angle changes, the coil is long enough to practically exclude the length influence on the results (Fig. 6). As one can see, 15% increase of the field can be reached when going from 45° to 30° tilt angle, which seems attractive, although tilt angle closer to 45$^O$ rather than to 30$^O$ can turn out to be more optimal from the practical point of view.

As can be also inferred from Fig. 6, the field dependence on the tilt angle is non-linear – roughly, the transverse field decreases directly with cosine of tilt angle. The same is observed for wire-wound tilted coils (see [3]). Thus, the use of very small tilt angles, 30$^O$ and less, is not so beneficial, especially if we also take into account that the decrease of tilt angle carries a penalty: more difficult assembling, considerable end effects and possibly lower mechanical strength.

## V. TILTED COIL HYBRID

As shown in [1], a highly uniform transverse field over 20T can be obtained with tilted Bitter coil magnet with a rather simple 3-coil configuration within 38 mm bore at reasonable power in the magnet. Unlike a split system, this field is uniform over the whole bore along a rather considerable portion of the magnet length (~100 mm and longer). Presumably, as estimates show, the maximum transverse field that can be reached with the present Florida-Bitter technology [1] within ~35-38 mm bore is about 25 T. In order to achieve 30 T, a hybrid configuration can be a good if not the only option with a Bitter insert and a superconducting outsert. For the latter, both a pancake technology described in [2] and layer-wise winding technology [3] can be used. Of course, quench analysis of superconducting coils is required to prove the feasibility. However, we are inclined to believe that quench behavior of tilted coils will not differ much from that of conventional superconducting magnets using similar conductors.

## VI. CONCLUSION

An elegant method is suggested to prove that a wire-wound tilted coil magnet has a generalized cosine-theta distribution of axial current density and, therefore, perfectly uniform transverse magnetic field over the whole bore. An exact solution for current density distribution within an elliptic disk of a tilted Bitter coil is given along with an efficient method of magnetic field calculation. Advantages and benefits from the use of tilted coil magnets are discussed.